\documentclass[12pt]{article}
\usepackage{amsfonts,amssymb,epsfig,amsmath}
\usepackage{slashed}
\usepackage{color}
\usepackage{hyperref}
\renewcommand{\text}[1]{#1}

\newcommand{\be}{\begin{equation}}

\newcommand{\ee}{\end{equation}}
\newcommand{\ben}{\begin{displaymath}}
\newcommand{\een}{\end{displaymath}}
\newcommand{\bea}{\begin{eqnarray}}
\newcommand{\eea}{\end{eqnarray}}
\newcommand{\bean}{\begin{eqnarray*}}
\newcommand{\eean}{\end{eqnarray*}}
\newcommand{\nn}{\nonumber \\}
\newcommand{\ba}{\begin{array}}
\newcommand{\ea}{\end{array}}
\newcommand{\bi}{\begin{itemize}}
\newcommand{\ei}{\end{itemize}}

\renewcommand{\theequation}{\arabic{section}.\arabic{equation}}
\def\theequation{\thesection.\arabic{equation}}

\def\a{\alpha}

\def\b{\beta}
\def\g{\gamma}
\def\G{\Gamma}

\def\G{\Gamma}
\def\g{\gamma}
\def\e{\epsilon}

\def\e{\epsilon}

\begin{document}

\makeatletter
\renewcommand{\theequation}{\thesection.\arabic{equation}}
\@addtoreset{equation}{section}
\makeatother

\begin{titlepage}

\vfill
\begin{flushright}
FPAUO-12/15\\
\end{flushright}

\vfill

\begin{center}
   \baselineskip=16pt
   {\Large \bf Fermionic T-duality: A snapshot review.}
   \vskip 2cm
    Eoin \'O Colg\'ain
       \vskip .6cm
             \begin{small}
      		 \textit{Departamento de F\'isica, 
		 Universidad de Oviedo, \\
33007 Oviedo, SPAIN}
             \end{small}\\*[.6cm]
\end{center}

\vfill \begin{center} \textbf{Abstract}\end{center} \begin{quote}
Through a self-dual mapping of the geometry $AdS_5 \times S^5$, fermionic T-duality provides a beautiful geometric interpretation of hidden symmetries for scattering amplitudes in $\mathcal{N} =4$ super-Yang-Mills. Starting with Green-Schwarz sigma-models, we consolidate developments in this area into this small review. In particular, we discuss the translation of fermionic T-duality into the supergravity fields via pure spinor formalism and show that a general class of fermionic transformations can be identified directly in the supergravity. In addition to discussing fermionic T-duality for the geometry $AdS_4 \times \mathbb{C} \textrm{P}^3$, dual to $\mathcal{N}=6$ ABJM theory, we review work on other self-dual geometries. Finally, we present a short round-up of studies with a formal interest in fermionic T-duality. 
\end{quote} \vfill

\end{titlepage}

\tableofcontents
%%%%%%%%%%%%%%%%%%%%%%%%%%%%%%%%%%%%%%%%%%%%%%%%%%%%%%%%%%%%%%%%%%%%%%%%%%%%%%%%%%%
%%%%%%%%%%%%%%%%%%%%%%%%%%%%%%%%%%%%%%%%%%%%%%%%%%%%%%%%%%%%%%%%%%%%%%%%%%%%%%%%%%%
\section{Introduction}
Given a non-linear sigma-model with a target space admitting an isometry, the Buscher procedure \cite{Buscher:1987sk} gives us a well-defined mechanism for gauging the isometry, integrating out the gauge fields and obtaining a T-dual sigma-model. The beauty of this approach is that beyond backgrounds admitting Abelian isometries, one may consider immediate generalisations to non-Abelian and fermionic isometries, leading to so called non-Abelian \cite{Buscher:1987sk, delaOssa:1992vc} and fermionic T-duality \cite{Berkovits:2008ic,Beisert:2008iq}. With many physical systems having non-Abelian symmetries, e.g. the strong nuclear force, Heisenberg ferromagnets and, at some scale, our homogeneous and isotropic universe, generalisations of Kramers-Wannier (Abelian) duality \cite{Kramers:1941kn} are certainly natural. In contrast,  the motivation to study fermionic generalisations has sprung out of a recent desire to explain striking hidden symmetries in scattering amplitudes in supersymmetric theories. 

These symmetries are surprising as they are not manifest in the initial Lagrangian formulation. Indeed, trading momentum for dual coordinates, it was first observed that perturbative MHV amplitudes exhibited an unexpected extra copy of the conformal symmetry, called ``dual conformal symmetry" \cite{Drummond:2006rz}, a symmetry that was later extended to the full superconformal symmetry \cite{Drummond:2008vq} in the context of $\mathcal{N}=4$ super-Yang-Mills (SYM). In a parallel development, gluon scattering amplitudes at strong-coupling were computed \cite{Alday:2007hr} and an astonishing connection between planar scattering amplitudes and Wilson loops in $\mathcal{N}=4$ SYM was elucidated \cite{Alday:2007he}. Complementary evidence for this connection also surfaced in weak-coupling computations \cite{Drummond:2007aua,Brandhuber:2007yx,Drummond:2007cf,Drummond:2007au}, and agreement at six-points and two-loops \cite{Drummond:2008aq,Bern:2008ap} is a noteworthy further robust test of this relationship. Later, a simple reformulation of scattering amplitudes  in terms of Grassmannians was proposed in  \cite{ArkaniHamed:2009dn,ArkaniHamed:2010kv}.

To account for the origin of  dual superconformal invariance and the duality between amplitudes and Wilson loops, ``fermionic T-duality" was introduced \cite{Berkovits:2008ic,Beisert:2008iq}. In fact, in the construction of classical string solutions AdS/CFT dual to gluon amplitudes \cite{Alday:2007hr, Alday:2007he} bosonic T-duality had already played a pivotal role in mapping $AdS_5$ back to itself. Building on this observation, \cite{Berkovits:2008ic,Beisert:2008iq} explain how the full superstring on $AdS_5 \times S^5$ can be mapped back to itself by considering additional novel fermionic transformations. In this process, the dual superconformal symmetry of scattering amplitudes maps to the ordinary superconformal symmetry of Wilson loops, and vice versa.  

Starting with the pioneering works \cite{Berkovits:2008ic,Beisert:2008iq}, in this review we present a snapshot picture of the current status of fermionic T-duality.  Following a brief summary of the Buscher procedure in section \ref{sec:Buscher}, in line with \cite{Berkovits:2008ic}, we discuss the natural generalisation to Green-Schwarz sigma-models, pure spinor sigma-models and the translation of the transformation into the fields of type II supergravity in section \ref{sec:fermionic}. From the supergravity perspective, we see that fermionic T-duality simply rotates Killing spinors, and thus, preserves both supersymmetry and chirality. 
Furthermore, if we require commuting fermionic isometries, we recognise that the transformation only allows non-Majorana Killing spinors, and as a result, complex solutions to supergravity typically ensue.  

Just as bosonic T-duality transformations can be derived without recourse to the sigma-model \cite{Bergshoeff:1995as}, in section \ref{sec:sugra} we summarise how a fermionic transformation generalising \cite{Berkovits:2008ic} may be identified through an educated ansatz  \cite{Godazgar:2010ph}. Formally, this allows \textit{real} transformations, and for massive IIA, the possibility of generating a Romans mass \cite{Romans:1985tz}. However, as we will discuss, in practice it appears that massive femionic T-duality is trivial and given a lower-dimensional D-brane in IIA, it is not possible to generate a mass term. 

In section \ref{sec:ads5s5} we discuss the exact self-duality of $AdS_5 \times S^5$ from the perspective of the both supergravity \cite{Berkovits:2008ic} and Green-Schwarz sigma-model \cite{Berkovits:2008ic, Beisert:2008iq}. We show that the dilaton shift characteristic of the bosonic T-dualities along $AdS_5$ is undone by fermionic T-dualities, leading to a symmetry at the quantum level. Furthermore, we explain that the family of flat currents,  or Lax connection, before and after T-dualities is related up to an automorphism of the superconformal algebra \cite{Beisert:2008iq} (see also \cite{Beisert:2009cs}).

The remainder of this review addresses generalisations, notably the search for self-dual geometries in $D \leq 10$ dimensions. In the light of mounting evidence in support of dual superconformal invariance in ABJM theory \cite{ABJM} at weak-coupling, we reflect on attempts to make this symmetry manifest at strong-coupling in the dual geometry $AdS_4 \times \mathbb{C} \textrm{P}^3$ \cite{Adam:2009kt, Grassi:2009yj, Adam:2010hh, Bakhmatov:2010fp}. Furthermore, we summarise recent work on a classification of fermionic T-duality symmetries of integrable Green-Schwarz sigma-models on AdS backgrounds with RR fluxes \cite{Dekel:2011qw}. The conclusion arrived at is that the only self-dual geometries are of the form $AdS_n \times S^n$, where $n =2,3, 5$. From the supergravity perspective, we comment on work on fermionic T-duality applied to D-branes \cite{Bakhmatov:2009be}, pp-waves \cite{Bakhmatov:2009be,Bakhmatov:2011aa} and present an additional example of an $AdS_n \times S^n$ ($n=3$) self-dual geometry embedded in $D=10$ \cite{OColgain:2012ca}. Finally, we provide a brief summary of work of a more formal flavour on fermionic T-duality. 

\section{T-duality \`a la Buscher} 
\label{sec:Buscher}
In this section we review the Buscher T-duality procedure \cite{Buscher:1987sk}. An earlier, more thorough review on bosonic T-duality can be found in \cite{Alvarez:1994dn}. We start by considering a sigma-model 
\be
S =  \int d^2 \zeta \left(  g_{ij}  + b_{ij} \right) \partial x^i \bar{\partial} x^j , 
\ee
where the symmetric tensor $g_{ij}$ and the antisymmetric tensor $b_{ij}$ denote the metric and $B$-field of the target space. 

Now, we assume that the target manifold admits an Abelian isometry. More concretely, we single out a particular direction, $x^1$, such that the background is invariant under constant shifts in $x^1$. This means that only derivatives of $x^1$ appear in the action. 

The Buscher procedure may be executed as follows. We replace derivatives of $x^1$ by a vector field $(A, \bar{A})$, while at the same time adding a Lagrange multiplier term $\tilde{x}^1$ to the action that ensures that the field strength vanishes. Doing so we have 
\bea
S &=&  \int d^2 \zeta \biggl[ g_{11} (x) A  \bar{A} +  l_{1 m} (x) A \bar{\partial} x^m   + l_{ m 1} (x) \partial x^m \bar{A} \nn && \phantom{xxxxxxxxxxxxxxxxxx} + l_{mn} \partial x^m \bar{\partial} x^n +  \tilde{x}^1 (\partial \bar{A} - \bar{\partial} A )  \biggr], 
\eea
where $l_{mn} = g_{mn} + b_{mn}$. Now, if we vary the action with respect to the Lagrange multiplier, we find the pure gauge condition $F = \partial \bar{A} - \bar{\partial} A =0$, which can be solved, $A = \partial x^1, \bar{A} = \bar{\partial} x^1$, and one recovers the original action. On the other hand, when $g_{11}$ is non-zero, we can integrate out the gauge field to get a T-dual sigma-model. Couplings in the T-dual sigma-model are then related to those in the original through the celebrated Buscher rules \cite{Buscher:1987sk}
\bea
\label{Buscher}
\tilde{g}_{11} &=& (g_{11})^{-1}, \quad \tilde{l}_{1 m} = (g_{11})^{-1} l_{1 m}, \quad \tilde{l}_{m 1} = -(g_{11})^{-1} l_{m 1}, \nn
\tilde{l}_{mn} &=& l_{mn} - (g_{11})^{-1} l_{m 1} l_{1 n}, \quad \phi' = \phi - \frac{1}{2} g_{11}.  
\eea

The accompanying shift in the dilaton is  the result of conveniently regularizing a measure factor coming from integration over the bosonic vector field. In turn, the regularization procedure is fixed by requiring conformal invariance of the dual theory \cite{Buscher:1987sk}. 

Some comments are now in order: 
\begin{enumerate}
\item In order for the transformation to be valid for an arbitrary compact Riemann surface we require that $x^1$ is compact. The reason being that on a arbitrary surface, $F=0$ does not imply that the gauge potential is exact. Instead, if we consider the theory on a sphere or a disk, we can still perform the transformation for non-compact scalars. 

\item The Buscher rules (\ref{Buscher}) generalised to include the transformation on the RR fluxes can also be derived directly from the supergravity. Starting from from either type IIA or type IIB supergravity, one can perform a circle reduction to get a unique $\mathcal{N}=2$ supergravity in $D=9$ \cite{Bergshoeff:1995as}. The transformation of the fields can then be read off from the lower-dimensional manifestation.  

\item The Buscher procedure naturally generalises for gaugings of non-Abelian isometries \cite{Buscher:1987sk,delaOssa:1992vc}. Recent progress means that given a type II supergravity solution, one can now generate a non-Abelian T-dual \cite{Sfetsos:2010uq,Lozano:2011kb,Itsios:2012dc} where the chirality of the resulting theory depends on the dimension of the gauge group.  In particular, if the dimension of the gauge group is even, chirality preserving transformations are possible. \end{enumerate}

\section{Fermionic T-duality} 
\label{sec:fermionic}
The Buscher procedure can naturally be extended to a Green-Schwarz type sigma-model depending on both bosonic and fermionic world sheet variables ($x^m, \theta^{\mu}$). This generalisation is succinctly described in \cite{Berkovits:2008ic}, so we simply present a short summary, but stress that the procedure mirrors the usual Buscher procedure, except now the isometry is fermionic. As before, we assume that the world sheet action is invariant under a constant shift, this time in one of the fermionic variables, which we denote $\theta^1$. This symmetry implies that only derivatives of $\theta^1$ appear in the action: 
\be
S = \int d^2 \zeta \left[ B_{11} \partial \theta^1 \bar{\partial} \theta^1 + L_{1M}  \partial \theta^1 \partial Y^{M} + L_{M1}  \partial Y^M \bar{\partial} \theta^1 + L_{MN} \partial Y^M \bar{\partial} Y^N\right], 
\ee
where the sigma-model coupling $L_{MN} = G_{MN}  + B_{MN} $ is the sum of the graded-symmetric tensor $G_{MN}$ and the graded-antisymmetric tensor $B_{MN}$. These are just functions of $Y^{M} = (x^m, \theta^{\mu})$, where $M = (m, \mu)$ ranges over all indices except for $\mu =1$.  

Now if $B_{11}$ is non-zero, one can apply the Buscher procedure as before. This is done by first introducing a fermionic vector field ($A, \bar{A}$), adding an accompanying Lagrange multiplier term to ensure that the vector field is the derivative of a scalar and then integrating out the fermionic gauge field to get the T-dualised action.  The T-dual sigma-model couplings are then related to the original couplings as follows \cite{Berkovits:2008ic}
\bea
\tilde{B}_{11} &=& - (B_{11})^{-1}, \quad \tilde{L}_{1M} = (B_{11})^{-1} L_{1M}, \quad \tilde{L}_{M1} = (B_{11})^{-1} L_{M1}, \nn
\tilde{L}_{MN} &=& L_{MN} - (B_{11})^{-1} L_{1N} L_{M1}, \quad   \tilde{\phi} = \phi + \frac{1}{2} \log B_{11}. 
\eea 
While expressions are analogous to the bosonic case, one striking departure is that the shift in the dilaton now comes with the opposite sign! This can easily be accounted for by the fact that we have integrated over an anticommuting variable and get an extra minus sign. So, the change in the dilaton $\phi$ under fermionic T-duality has the opposite sign from the change   under bosonic T-duality, a fact that will be important later when we consider the self-duality of the geometry $AdS_5 \times S^5$.  

Furthermore, throughout this process, our fermionic variables have been taken to be non-compact. Thus, this derivation only holds in the case of the disk but not on higher genus Riemann surfaces. That fermionic T-duality is not a full symmetry of String theory is apparent from how S-duality affects the self-duality properties of geometries \cite{OColgain:2012ca}.

As mentioned in passing earlier, two-derivative terms such as $\int d^2 \zeta B_{11} \partial \theta^1 \bar{\partial} \theta^1$ typically arise in Green-Schwarz sigma-models for type II superstrings in Ramond-Ramond backgrounds. So, given the sigma-model description, one can perform the fermonic T-duality transformation as sketched above and read off the T-dual background. Ideally, one could perform fermionic T-duality transformations directly on the type II supergravity background fields. This indeed can be done as pointed out in \cite{Berkovits:2008ic}. To do this, we need to understand the relationship of the sigma-model couplings to the on shell supergravity fields. For both bosonic \cite{Benichou:2008it} and fermionic T-duality a convenient method for establishing this connection is to use pure spinor formalism where BRST invariance determines the choice of torsion conditions and facilitates the identification of the background fields. The fermionic T-duality transformation rules we will recover later by separate means. 

For simplicity, we focus on the relevant fields and refer the reader to \cite{Berkovits:2008ic} for more details. In the pure spinor version of the type II sigma model, the worldsheet action is 
\be
S = \int d^2 \zeta \left[ L_{MN} \partial Z^M \bar{\partial} Z^N + P^{\a \hat{\beta}} d_{\alpha} \hat{d}_{\hat{\beta}} + E_{M}^{\alpha} d_{\alpha} \bar{\partial} Z^M + E^{\hat{\alpha}}_{M} \partial Z^M \hat{d}_{\hat{\alpha}}  + \dots \right], 
\ee
where $Z^M$ are coordinates for the $\mathcal{N}=2, D=10$ superspace, $d_{\a}, \hat{d}_{\hat{\a}}$ are independent fermionic variables and dots denote terms we have dropped for brevity. BRST invariance implies relations between various superfields appearing in the action that we will address soon. Of present particular interest  is the $\theta = \hat{\theta} = 0$ component of $P^{\a \hat{\b}}$ and $(E_{1}^{\a}, E_1^{\hat{\a}} )$ which correspond to the RR flux bispinor and the Majorana-Weyl Killing spinors of type II supergravity, respectively: 
\bea
P^{\a \hat{\beta}}|_{\theta = \hat{\theta} =0} &=& - \frac{i}{4} e^{\phi} F^{\a \hat{\b}}, \nn
&=& - \frac{i}{4}  e^{\phi} \left[ (\g^m)^{\a \hat{\b}} F^{(1)}_{m}+ \frac{1}{3!} (\g^{mnp})^{\a \hat{\b}} F_{mnp}^{(3)}+ \frac{1}{2 (5!)} ( \g^{mnpqr})^{\a \hat{\b}} F^{(5)}_{mnpqr}\right], \nn
(E^{\a}_{1}, E^{\hat{\a}}_{1}) |_{\theta = \hat{\theta} =0}  &=& (\e^{\a}, \hat{\e}^{\a}).  
\eea
We remark that only the transformation rules in the context of type IIB supergravity are discussed in \cite{Berkovits:2008ic}, so only the odd RR-forms appear above. However, it is expected that this procedure will apply equally well to type IIA. Furthermore, we should point out that as our Killing spinors ($\e^{\a}, \hat{\e}^{\hat{a}}$) are assumed to be Majorana-Weyl, our gamma matrices here are $16 \times 16$ matrices. 

Then following the T-duality prescription of the last subsection, one finds that the superfields highlighted above transform as
\be
\label{superf}
\tilde{P}^{\a \hat{\b}} = {P}^{\a \hat{\b}} - (B_{11})^{-1} E_1^{\a} E_1^{\hat{\b}}, \quad \tilde{E}_{1}^{\a} = (B_{11})^{-1} E_1^{\a}, \quad \tilde{E}_{1}^{\hat{\a}} = (B_{11})^{-1} E_1^{\hat{\a}}.
\ee

At this stage we can comment on one important distinction between bosonic and fermionic T-duality. While for bosonic T-duality, a relative sign appears\footnote{Bosonic T-duality on the $x^p$ coordinate gives $ \tilde{E}_{p}^{\a} = (G_{pp})^{-1} E_{p}^{\a}, ~~ \tilde{E}_{p}^{\hat{\a}} = -(G_{pp})^{-1} E_{p}^{\hat{\a}}$ \cite{Benichou:2008it}. }, above $\tilde{E}_{1}^{\a}$ has the same sign as $\tilde{E}^{\hat{\a}}_1$. It is known \cite{Berkovits:2001ue} that BRST invariance of the sigma model implies the superspace torsion constraints 
\be
T^{a}_{\a \b} = i f^{a}_{b} \g^{b}_{\a \b}, \quad T^a_{\hat{\a} \hat{\b}} = i \hat{f}^{a}_{b} \g^{b}_{\hat{\a} \hat{\b}}, 
\ee
where $f^{a}_{b}$ and $\hat{f}^{a}_{b}$ are $O(9,1)$ matrices, which in turn should be gauge fixed to $f^{a}_{b} = \hat{f}^{a}_{b} = \delta^{a}_{b}$ to recover the torsion constraints of type II supergravity. Then, performing a bosonic T-duality means that a relative sign is introduced between $f^{a}_{b}$ and $\hat{f}^{a}_{b}$, so to recover the original torsion conditions one has to perform a local Lorentz transformation and flip the chirality of either the hatted or unhatted spinors. Such a change of chirality under fermionic T-duality is not required. Further discussion on this point appears in \cite{Berkovits:2008ic}. 

We are now in a position to say something about the transformation rules. By considering the $\theta = \hat{\theta}=0$ components of the superfields (\ref{superf}) one finds that the RR bispinor transforms as follows: 
\bea
- \frac{i}{4} e^{\tilde{\phi}} \tilde{F}^{\a \hat{\beta}} = - \frac{i}{4} e^{\phi} F^{\a \hat{\b}} - \epsilon^{\a} \hat{\epsilon}^{\hat{\beta}} C^{-1}, ~~\textrm{where} ~~ C = B_{11} |_{\theta = \hat{\theta} = 0} . 
\eea
From expressions we have omitted above (see \cite{Berkovits:2008ic} for more details), one can see that the NS fields $g_{mn}$ and $b_{mn}$ do not change, and as before, we have the same shift in the dilaton, $\tilde{\phi} = \phi + \frac{1}{2} \log C$.  

To complete the picture, we now need to identify the relation between $C$ and the supergravity Killing spinors $(\e^{\a}, \hat{\e}^{\hat{\a}})$. To do this one can use the fact \cite{Berkovits:2008ic} that the torsion constraints imply that the superspace 3-form field strength 
\be
H_{ABC} = E_{A}^{M} E_{B}^{N} E^{P}_{C} \partial_{[M} B_{NP]} 
\ee
has constant spinor-spinor-vector components \cite{Howe:1983sra}
\be
H_{\a \b c} = i (\g_c)_{\a \b}, \quad H_{\hat{\a} \hat{\b} c} = - i (\g_c)_{\hat{\a} \hat{\b}}, \quad H_{\a \hat{\b} c} = 0. 
\ee
Here $A = (c, \a, \hat{\a})$ denotes tangent-superspace indices, $M$ denotes curved-superspace indices, and $E_{A}^M$ is the inverse super-vielbein. Since we have a fermionic isometry, we know $\partial_{\theta^1} B_{1m} =0$, allowing us to determine $C$ in terms of $(\e^{\a}, \hat{\e}^{\hat{\a}})$
\bea
\partial_{m} C &=& \partial_{m} B_{11} |_{\theta = \hat{\theta} = 0} = E_{1}^{A} E_{1}^{B} E_{m}^{C} H_{ABC} |_{\theta = \hat{\theta} = 0}, \nn
&=& i \e^{\a} \e^{\b} e^{c}_{m} (\g_c)_{\a \b} - i \hat{\e}^{\hat{\a}} {\hat{\e}}^{\b} e^{c}_{m} (\g_c)_{\hat{\a} \hat{\b}}, \nn
&=& i (\e \g_m \e - \hat{\e} \g_m \hat{\e}), 
\eea
where $e^{c}_{m} = E^{c}_{m} |_{\theta = \hat{\theta} =0}$ is the usual vielbein and we have suppressed indices in the last line. 

So, to summarise what we have established so far; type II supergravity admits a fermionic symmetry defined by a pair of Killing spinors ($\e, \hat{\e}$) which together correspond to a fermionic  isometry direction. Given the Killing spinors,  one can determine $C$, which started out as the $\theta = \hat{\theta} =0$ component of the sigma-model coupling $B_{11}$, and the transformations of RR fluxes and the dilaton follow from a knowledge of $C$. However, the Killing spinors cannot be chosen randomly \cite{Berkovits:2008ic}. Since the fermionic isometry is assumed to be Abelian, we have $\{ \e^{\a} Q_{\a}, \hat{\e}^{\hat{\a}} Q_{\hat{\a}} \} =0 $, which can only be consistent with the supersymmetry algebra $\{ \e^{\a} Q_{\a}, \hat{\e}^{\hat{\a}} Q_{\hat{\a}} \} = (\e \g^{m} \e + \hat{\e} \g^m \hat{\e} ) P_{m}$ if 
\be
\label{compspinor}
\e \g^m \e + \hat{\e} \g^m \hat{\e} = 0. 
\ee
Note that if ($\e, \hat{\e}$) are Majorana spinors, as they usually are,  it is not possible to satisfy this condition, so the only non-trivial solutions involve complex Killing spinors. An immediate drawback then is that fermionic T-duality generically produces complex solutions to supergravity, unless one compensates by performing, for example, a timelike T-duality \cite{Hull:1998vg}. 

So far we have assumed that we are doing a single fermionic T-duality, but all formulae above can be immediately extended to transformations involving $I, J= 1,\dots n$ commuting fermionic isometries. In this case the transformation rules become:
\bea
\label{commutef} 0 &=& \e_{I} \g^m \e_{J} + \hat{\e}_{I} \g^m \hat{\e}_{J} , \\
\label{diffC} \partial_{m} C_{IJ} &=& 2 i \e_{I} \g_m \e_{J},  \\
\label{dil1} \tilde{\phi} &=& \phi + \frac{1}{2} \textrm{Tr} (\log C), \\
\label{fluxtrans} - \frac{i}{4} e^{\tilde{\phi}} \tilde{F}^{\a \hat{\beta}} &=& - \frac{i}{4} e^{\phi} F^{\a \hat{\b}} - \epsilon^{\a}_{I} \hat{\epsilon}^{\hat{\beta}}_{J} (C^{-1})_{IJ}, 
\eea
As a closing comment in this section we remark that generically fermionic T-duality preserves the number of supersymmetries. Explicitly the Killing spinors of the T-dual theory are 
\be
\tilde{\e}^{\a}_{I} = (C^{-1})_{IJ} \e^{\a}_{J}, \quad  \tilde{\hat{\e}}^{\hat{\a}}_{I} = (C^{-1})_{IJ} \hat{\e}^{\hat{a}}_{J}.  
\ee
%\subsection{Supersymmetry of T-dualised background} 

\section{Further insights from supergravity}
\label{sec:sugra}
In much the same way as bosonic T-duality can be understood in terms of the low-energy effective action \cite{Bergshoeff:1995as}, the fermionic T-duality transformation rules can also be worked out without having to resort to pure spinor formalism. This approach was adopted in \cite{Godazgar:2010ph}, and not only does it offer an independent derivation of the fermionic T-duality rules for both type IIB and type IIA supergravity, but also presents a generalisation which formally allows \textit{real} transformations\footnote{While the transformation rules exist, no examples are known.}.

%Recall from earlier that the formulation of Berkovits \& Maldacena means that fermionic T-duality will generically produce complex supergravity backgrounds unless combined with a timelike bosonic T-duality. 

In this section, following \cite{Godazgar:2010ph}, we give an account of how the fermionic T-duality rules for type IIA supergravity are derived. The procedure for type IIB runs along similar lines, so we omit the details and recommend the interested reader to \cite{Godazgar:2010ph}.  In some sense the type IIA analysis is more interesting as it can be generalised to massive IIA supergravity \cite{Romans:1985tz}, a point which we will return to in due course. For the moment we confine our attention to the massless case. We start by considering an ansatz for the transformation of the RR fluxes and the dilaton of the form \cite{Godazgar:2010ph}: 
\bea
\label{RRansatz}
e^{\phi} F_{ab}^{(2)} &\mapsto& e^{\tilde{\phi}} \tilde{F}^{(2)}_{ab} = e^{\phi} F^{(2)}_{ab} + \bar{\e}_{I} \G_{ab} (S_1 + S_2 \G_{11}) \eta_{J} M_{IJ}, \nn
e^{\phi} F_{abcd}^{(4)} &\mapsto& e^{\tilde{\phi}} \tilde{F}^{(4)}_{abcd} = e^{\phi} F^{(4)}_{abcd} + \bar{\e}_{I} \G_{abcd} (S_3 + S_4 \G_{11}) \eta_{J} M_{IJ}, \nn
\tilde{\phi} &=& \phi +X,
\eea
where $X$, $M_{IJ}$ and $S_i$, $i=1,\dots,4$ are arbitrary functions to be determined and the spinors $\e_{I}, \eta_{I}$ correspond to real solutions to the type IIA Killing spinor equations
\bea
\label{grav} \left[ \nabla_{a}  - \frac{1}{8} H_{abc} \G^{bc} \G_{11}  - \frac{1}{16} e^{\phi} F^{(2)}_{bc} \G^{bc} \G_{a} \G_{11} + \frac{1}{192} e^{\phi} F^{(4)}_{bcde} \G^{bcde} \G_{a} \right]  \e &=& 0, \\
\label{dilatino} \left[  \G^{a} \partial_{a} \phi - \frac{1}{12} H_{abc} \G^{abc} \G_{11} - \frac{3}{8} e^{\phi} F^{(2)}_{ab} \G^{ab} \G_{11} + \frac{1}{96} e^{\phi} F^{(4)}_{abcd} \G^{abcd} \right] \e &=& 0. 
\eea
Throughout this section our gamma matrices are $32 \times 32$ matrices, so to highlight this distinction we have adopted different notation. Furthermore, our Killing spinor is now the sum of two Majorana-Weyl spinors. 

Now, observe that on the RHS of (\ref{RRansatz}) we have shifted the RR fluxes by Killing spinor bilinears contracted into a matrix $M_{IJ}$, so the LHS is \textit{a priori} not a solution to the equations of motion of IIA supergravity. To remedy this we will now impose the Bianchi identity, the flux equations of motion, the dilaton equation and lastly the Einstein equation. 

As an example, we focus on the Bianchi $d \tilde{F}^{(2)} = 0$. To impose this condition we utilise the gravitino variation (\ref{grav}) to differentiate the Killing spinor bilinear and the dilatino variation (\ref{dilatino}) to simplify the resulting expressions. This expression must then vanish for the transformed Bianchi identity to be satisfied. If in addition, one assumes that one is considering generic supergravity solutions, one can require that terms proportional to the RR two-form $F^{(2)}$, the NS field strength $H=d B$, the RR four-form $F^{(4)}$ and the remaining terms vanish independently. In the process one derives a set of differential conditions and constraints. One then proceeds to the other equations of motion in turn and in each case one records the constraints. After considerable work, occasionally involving the Fierz identity, one finds a series of conditions that may be solved by the following constraints \cite{Godazgar:2010ph}:
% \cite{Godazgar:2010ph}
%\bea
%\partial_{a} X &=& S_1 \bar{\e}_{I} \G_{a} \G_{11} \eta_{J} M_{IJ}, 
%\eea
%and constraints 
%\bea
%0 &=& S_1 \bar{\e}_{I} \G_{abc} \eta_J M_{IJ} =  \bar{\e}_{I} \left(  S_1 \G_{11} + S_2  \right) \eta_J M_{IJ}, \nn
%0 &=& S_1 \bar{\e}_{I} \G_{a} \eta_J M_{IJ} = S_2  \bar{\e}_{I} \G_{abc} \G_{11} \eta_J M_{IJ}, \nn
%0 &=& \bar{\e}_{I} \G_{[bc} \left( \partial_{a]} X (S_1 + S_2 \G_{11} ) M_{IJ} - \partial_{a]} [(S_1 + S_2 \G_{11}) M_{IJ} ]\right) \eta_J. 
%\eea
\bea
\label{const1} 0 &=& \bar{\e}_{I} \G_{a} \eta_J \\ 
\label{const2} 0 &=& \bar{\e}_I \G_{11} \eta_J M_{IJ}, \\
\label{const3} 0 &=& \bar{\e}_I \eta_J M_{IJ}  = \bar{\e}_{I} \G_{ab} \G_{11} \eta_J M_{IJ}, \\
\label{const4} 0 &=& \bar{\e}_{I} \G_{abc} \eta_J M_{IJ} = \bar{\e}_{I} \G_{abc} \G_{11} \eta_J M_{IJ} = \bar{\e}_{I} \G_{abcd} \eta_{J} M_{IJ}, \\
\label{diffX} \partial_{a} X &=& \bar{\e}_{I} \G_a \G_{11} \eta_J M_{IJ}, \\
\label{diffM} \partial_{a} M_{IJ} &=& - 2 \bar{\e}_{K} \G_{a} \G_{11} \eta_L M_{IL} M_{KJ}. 
\eea
Note, this is the only solution found by the authors \cite{Godazgar:2010ph}, but this may not be exhaustive. 

Then, provided the Killing spinors satisfy these conditions, the type IIA supergravity equations admit the following symmetry: 
\bea
\phi  &\mapsto& \tilde{\phi} = \phi + X, \nn
e^{\phi} F_{ab}^{(2)} &\mapsto& e^{\tilde{\phi}} \tilde{F}^{(2)}_{ab} = e^{\phi} F^{(2)}_{ab} + \bar{\e}_{I} \G_{ab} \eta_{J} M_{IJ}, \nn
e^{\phi} F_{abcd}^{(4)} &\mapsto& e^{\tilde{\phi}} \tilde{F}^{(4)}_{abcd} = e^{\phi} F^{(4)}_{abcd} - \bar{\e}_{I} \G_{abcd} \G_{11} \eta_{J} M_{IJ}. 
\eea
Observe that this corresponds to the original ansatz (\ref{RRansatz}) with $S_1 = -S_4 =1$ and $S_2 = S_3 =0$. Also observe that (\ref{diffM}) may be rewritten as 
\be
\partial_a (M^{-1})_{IJ} = 2 \bar{\e}_I \G_{a} \G_{11} \eta_J, 
\ee
whereas (\ref{diffX}) can be solved up to an integration constant 
\be
X = \frac{1}{2} \textrm{Tr} (\log M^{-1}), 
\ee
both of which are reminiscent of the fermionic T-duality transformation of Berkovits \& Maldacena once one identifies the inverse of the matrix $M$ with $C$. 

Certainly, as a special case, this symmetry includes the Fermionic T-duality of  \cite{Berkovits:2008ic}. To see this we set $\eta_{I} = \e_{I}$ in which case only the symmetric part of $M_{IJ}$ plays a role when contracted into the spinor bilinears. Then, as an immediate consequence of the symmetry properties of the gamma matrices it can be shown that the constraints (\ref{const3}) and (\ref{const4}) are now trivially satisfied.  (\ref{const1}) is the familiar constraint that comes from the assuming the fermionic isometries are Abelian (\ref{commutef}). As an added bonus, this constraint can also be seen as a direct consequence of the integrability conditions arising from (\ref{diffX}) and (\ref{diffM}) \cite{Godazgar:2010ph}. Finally, the constraint (\ref{const2}) may be physically interpreted as maintaining a zero Romans mass \cite{Romans:1985tz}, so it is valid to wonder if one can generate a mass. 

In a formal sense, this is possible. One can extend the assumed ansatz to include a shift in the mass term  \cite{work}
\be
e^{\phi} m \mapsto e^{\tilde{\phi}} \tilde{m} = e^{\phi} m + \bar{\e}_{I} \G_{11} \eta_{J} M_{IJ}. 
\ee
Then the problem boils down to finding solutions where the spinor bilinear on the RHS is non-zero. When $\eta_{I} = \epsilon_{I}$ it is possible to show that if one tries to apply fermionic T-duality to characteristic solutions of massive IIA, such as D8-branes and the warped product $AdS_6 \times S^4$ \cite{Brandhuber:1999np}, the constraint equation implies $M_{IJ}$ is a constant and that the transformation is trivial. So, the picture emerging is that while it is a formal symmetry of massive type IIA, fermionic T-duality transformations on standard massive solutions are trivial. From a separate perspective, one can ask whether one can generate a mass term from lower-dimensional Dp-branes. As fermionic T-duality does not change the metric, it is unlikely that one can convert a D2-brane directly into a D8-brane, but one may ask can a mass be generated? By looking at the Killing spinor projection conditions for various Dp-branes, one can confirm that the bilinear $\bar{\e}_{I} \G_{11} \eta_{J}$ has to be zero, so this possibility also appears remote.

\section{Exact T-duality of $AdS_5 \times S^5$}
\label{sec:ads5s5}
In this section we show that the geometry $AdS_5 \times S^5$ is self-dual with respect to a particular combination of bosonic and fermionic T-dualities. We will focus on bosonic T-duality with respect to the four coordinates $(x^0, x^1, x^2, x^3)$ of $AdS_5$ in Poincar\'e coordinates and eight fermionic T-dualities involving Poincar\'e supersymmetries \cite{Berkovits:2008ic, Beisert:2008iq}. As noted in \cite{Berkovits:2008ic}, a second possibility also exists where, if one breaks the $SU(4)$ R-symmetry to $U(1) \times SU(2) \times SU(2)$, one can choose eight Abelian Poincar\'e supersymmetries with charge $+1$ with respect to the $U(1)$ direction. To establish self-duality, this combination requires additional internal bosonic T-dualities with respect to four directions corresponding to $SU(4)$ generators with charge $+2$ under the $U(1)$ \cite{Berkovits:2008ic}. 

Before discussing the T-duality for the $AdS_5 \times S^5$ sigma-model \cite{Metsaev:1998it}, we sketch the supergravity calculation. Employing Poincar\'e coordinates for $AdS_5$
\be
ds^2(AdS_5) = r^{-2} \left( dx^m dx_m + dr^2 \right), 
\ee
where $m = 0, 1, 2, 3$, we can solve for the Poincar\'e Killing spinors $\eta$ of $AdS_5 \times S^5$ to find  $\eta = r^{-1/2} \tilde{\eta}$, where we have absorbed all angular dependence on the $S^5$ in $\tilde{\eta}$. The geometry $AdS_5 \times S^5$ is supported by a self-dual five-form flux with a trivial dilaton, meaning that, before we perform any fermionic T-duality, the flux bispinor is simply 
\be
\label{origspinor}
 e^{{\phi} } {F}^{\a \hat{\b}} = (\g_{01234})^{\a \hat{\b}},  
\ee 
where $0,1,2,3$ correspond to the $x^m$ directions and $4$ is the $AdS_5$ radial direction. 
As we have sixteen original Poincar\'e supersymmetries, the effect of the constraint (\ref{commutef}) is to pick out eight pairs of Killing spinors corresponding to eight commuting fermionic directions. Proceeding, one can determine $C_{IJ}$ from (\ref{diffC}), invert and contract into the Killing spinors. Adding this matrix to our original RR flux bispinor (\ref{origspinor}) as guided by (\ref{fluxtrans}), one finds the fermionic T-dual bispinor \cite{Berkovits:2008ic}
\be
e^{\tilde{\phi}} \tilde{F}^{\a \hat{\b}} = (i \g_4)^{\a \hat{\b}}. 
\ee
Note that the effect of the eight commuting fermionic T-dualities is to replace our original five-form flux with a one-form flux proportional to $dr$, which is now \textit{complex}. The dilaton shift can be worked out from (\ref{dil1}) and one discovers that $\phi$ is shifted as follows
\be
\tilde{\phi} = \phi + 4 \log r. 
\ee
At this stage, to recover the original geometry, we simply have to perform bosonic T-dualities along $x^m$. In the process the shift in the dilaton is undone, since from (\ref{Buscher}) we see that $\delta \phi = -4 \log r$,  and the timelike T-duality results in all RR fluxes picking up an $i$ factor \cite{Hull:1998vg}, so that we recover the real solution with which we started. 

\subsection{Green-Schwarz sigma-model}
We now turn our attention to describing the type IIB Green-Schwarz superstring action on $AdS_5 \times S^5$. So far we have followed mostly \cite{Berkovits:2008ic}, so here we opt to present the discussion in \cite{Beisert:2008iq}. The conclusions reached by \cite{Berkovits:2008ic} are the same. We will see that a series of bosonic and fermionic T-dualities map the action back to itself. 

The $AdS_5 \times S^5$ superstring action can be understood as a sigma-model type action on the coset superspace $G/H = PSU(2,2|4) / (SO(1,4) \times SO(5) )$ \cite{Metsaev:1998it}, where the bosonic part simply corresponds to the isometries of the geometry. The coset admits a $\mathbb{Z}_4$-grading \cite{Berkovits:1999zq}, which means that at the level of the Lie algebra $\mathfrak{g}:= \textrm{Lie}(G)$, one has 
\be
\mathfrak{g} \cong \bigoplus_{m=0}^{3} \mathfrak{g}_{(m)}, ~~~\textrm{with}~~\mathfrak{g}_{(0)} := \textrm{Lie} (H) ~~\textrm{and} ~~ [\mathfrak{g}_{(m)}, \mathfrak{g}_{(n)}] \subset \mathfrak{g}_{(m+n)}.  
\ee 

To define the superstring action, one considers the map $g: \Sigma \rightarrow G$, where $\Sigma$ is the string worldsheet and introduces the current 
\be
\label{current}
j = g^{-1} d g = j_{(0)} + j_{(1)} + j_{(2)} + j_{(3)}, ~~\textrm{with}~~ j_{(m)} \in \mathfrak{g}_{(m)}. 
\ee 
We can then write down a superstring action \cite{Metsaev:1998it,Berkovits:1999zq,Roiban:2000yy}
\be
\label{act}
S = - \tfrac{T}{2} \int_{\Sigma} \textrm{str} \left[ j_{(2)} \wedge * j_{(2)} + \kappa j_{(1)} \wedge j_{(3)}\right], 
\ee
where $T$ is the string tension, $*$ is the worldsheet Hodge star and `$\textrm{str}$' denotes the supertrace. The parameter $\kappa$ is set through a $\kappa$-symmetry condition requiring $\kappa = \pm 1$. In what follows it is assumed that $\kappa =1$. 

By construction, the action is invariant under a global (left) $G$-transformation and local (right) $H$-transformation of the form 
\bea 
\label{left} g \mapsto g_0 g, ~~ g_0 \in G \\ \label{right} g \mapsto g h, ~~h \in H. 
\eea
Indeed, the current $j$ is invariant under (\ref{left}), while under (\ref{right}) $j_{(0)}$ transforms as a connection, $j_{(0)} \mapsto h^{-1} j_{(0)} h + h^{-1} d h$ and $j_{(m)}$, $m=1,2,3$ transform covariantly, $j_{(m)} \mapsto h^{-1} j_{(m)} h$. 

More concretely, to construct an explicit action, we adopt a specific choice for the basis generators of $\mathfrak{g} = \mathfrak{psu}(2,2|4)$  \cite{Beisert:2008iq}:
\be
\mathfrak{psu}(2,2|4) = \textrm{span} \{P_{a}, L_{ab}. K_{a}, D, R_{i}^{~j} | Q^{i \a}, \bar{Q}_{i}^{\dot{\a}}, S_{i}^{\a}, \bar{S}^{i \dot{\a}} \}, 
\ee
where essentially, $P, L, K, D$ denote the generators of the conformal group, $R$ corresponds to the $SU(4)$ R-symmetry generators and $Q, \bar{Q}$, $S, \bar{S}$ denote Poincar\'e and superconformal supercharges and their Hermitian conjugates.  In terms of these generators, the $\mathbb{Z}_4$-splitting above may be expressed as \cite{Beisert:2008iq}
\bea
\mathfrak{g}_{(0)} &=& \textrm{span} \{ \tfrac{1}{2} (P_a - K_a), L_{ab}, R_{(ij)} \}, \nn
\mathfrak{g}_{(1)} &=& \textrm{span} \{ \tfrac{1}{2} (Q^{i \a} + C^{ij} S_{j}^{\a}), \tfrac{1}{2} ( \bar{Q}^{\dot{\a}}_{i} + C_{ij} \bar{S}^{j \dot{\a}} ) \}, \nn
\mathfrak{g}_{(2)} &=& \textrm{span} \{ \tfrac{1}{2} (P_a + K_a), D, R_{[ij]} \}, \nn
\mathfrak{g}_{(3)} &=& \textrm{span} \{ -\tfrac{i}{2} (Q^{i \a} - C^{ij} S_{j}^{\a}), \tfrac{i}{2} ( \bar{Q}^{\dot{\a}}_{i} - C_{ij} \bar{S}^{j \dot{\a}} ) \}, 
\eea
where $R_{ij} = C_{ik} R_{j}^{~k}$ and the constant matrix $C_{ij}$ is an $Sp(4)$-metric that may be interpreted as charge conjugation. Note that the $\mathbb{Z}_4$-gradation of the Lie algebra means that $\mathfrak{g}_{(0)}$ and $\mathfrak{g}_{(2)}$  are generated by the bosonic operators, while $\mathfrak{g}_{(1)}$ and $\mathfrak{g}_{(3)}$ are generated by fermionic ones. 

As a next step, one would like to find an explicit form for the current $j$ of the action (\ref{act}) tailored to the Poincar\'e form of the $AdS_5 \times S^5$ metric
\be
\label{metric}
ds^2 = - \tfrac{1}{2} Y^2 d X_{\a \dot{\b}} d X^{\dot{\b} \a} + \tfrac{1}{4 Y^2} d Y_{ij} d Y^{ij}, 
\ee
where, without going into details of the notation (appendix A of \cite{Beisert:2008iq}), $X_{\a \dot{\b}}$ label the coordinates on $\mathbb{R}^{1,3}$ and $Y_{ij}$ denote coordinates on $\mathbb{R}^6$, including the $AdS_5$ radial direction, $Y^2 = \frac{1}{4} Y_{ij} Y^{ij}$.  We can find a simplified form for $j$ by appropriately fixing the local $H$-symmetry  \cite{Metsaev:2000yf,Metsaev:2000yu} so that $g$ is of the form 
\be
g (X, Y, \Theta) = B(X,Y) e^{-F(\Theta)}, 
\ee
where $\Theta = (\theta^{\a i}_{\pm}, \bar{\theta}^{\dot{\a}}_{\pm i} )$ denote  32 independent fermionic coordinates obeying a reality condition:
\be
\label{real}
\theta^{i \a}_{\pm} = (\bar{\theta}^{\dot{\a}}_{\pm i} )^{\dagger}. 
\ee
We can further simplify the expression for $j$ by adopting a specific $\kappa$-symmetry gauge choice so that only the Poincar\'e supercharges appear, i.e. by setting $\theta^{i \a}_{-} = \bar{\theta}^{\dot{\a}}_{-i} =0$. 

Doing so, one can determine $j$ and extract out $j_{(m)}$, $m =1,2,3$ so that the action takes the form \cite{Beisert:2008iq}
\bea
\label{origact}
&&S = - \tfrac{T}{2} \int_{\Sigma} \biggl\{ - \tfrac{1}{2} Z^2 \Pi_{\a \dot{\b}} \wedge * \Pi^{\dot{\beta} \a} + \tfrac{1}{4 Z^2} d Z_{ij} \wedge * d Z^{ij} \nn && ~~~~~~~~~~~~~~~~~~~~+ \tfrac{1}{2} \left( d Z_{ij} \wedge \theta^{i \a} d \theta^{j }_{\a} - d Z^{ij} \wedge \bar{\theta}^{\dot{\a}}_{i} d \bar{\theta}_{j \dot{\a} }\right) \biggr\},  
\eea
where $Z_{ij}$ are related to the earlier $Y_{ij}$ by a coordinate transformation \cite{Beisert:2008iq} and we have defined 
\be
\Pi^{\a \dot{\b}} = d X^{\dot{\a} \b} + \tfrac{i}{2} ( \bar{\theta}^{\dot{\a}}_{i} d \theta^{i \beta} - d \bar{\theta}_i^{\dot{\a}} \theta^{i \beta}). 
\ee
 
We can now perform the Buscher procedure on this action by introducing an auxiliary one-form $V$ and a field $\tilde{X}^{\a \dot{\b}}$, which plays the role of a Lagrange multiplier ensuring that $V$ is flat, i.e. $d V = 0 \Rightarrow V = d X$. 

On the other hand, solving for $V$, we have  
\be
V^{\dot{\a} \b} + \tfrac{i}{2} ( \bar{\theta}^{\dot{\a}}_{i} d \theta^{i \beta} - d \bar{\theta}^{\dot{\a}}_i \theta^{i \beta} ) = Z^{-2} * d \tilde{X}^{\dot{\a} \b}, 
\ee
and after substituting in for $V$ in the action, we find the resulting T-dual action in terms of the T-dual coordinate $\tilde{X}^{\a \dot{\b}}$ \cite{Beisert:2008iq}
\bea
\label{tdualact} 
&&S =  - \tfrac{T}{2} \int_{\Sigma} \biggl\{ - \tfrac{1}{2 Z^2}   d \tilde{X}_{\a \dot{\b}} \wedge * d \tilde{X}^{\dot{\beta} \a} + \tfrac{1}{4 Z^2} d Z_{ij} \wedge * d Z^{ij} \nn && + \tfrac{i}{2} d \tilde{X}_{\b \dot{\a}} \wedge (\bar{\theta}^{\dot{\a}}_{i} d \theta^{i \b} - d \bar{\theta}^{\dot{\a}}_{i} \theta^{i \beta}) + \tfrac{1}{2} \left( d Z_{ij} \wedge \theta^{i \a} d \theta^{j }_{\a} - d Z^{ij} \wedge \bar{\theta}^{\dot{\a}}_{i} d \bar{\theta}_{j \dot{\a} }\right) \biggr\}.  
\eea
One can observe that the geometry is still $AdS_5 \times S^5$, since one can change the coordinates $Z_{ij}$ so that $Z \mapsto Z^{-1}$ and recover the original form for the bosonic part of the action. 

Now, as explained further in \cite{Beisert:2008iq}, starting from (\ref{tdualact}), we can perform further fermionic T-duality transformations on the coordinates $\theta^{i \a}$ (but not their conjugates $\bar{\theta}^{\dot{\a}}_{i}$) and find the fermionic T-dual action:
\bea
\label{ftdualact} 
&&S =  - \tfrac{T}{2} \int_{\Sigma} \biggl\{ - \tfrac{1}{2 Z^2}   d \tilde{X}_{\a \dot{\b}} \wedge * d \tilde{X}^{\dot{\beta} \a} + \tfrac{1}{4 Z^2} d Z_{ij} \wedge * d Z^{ij} \nn && - \tfrac{1}{2 Z^2} Z^{ij} \e^{\a \b} ( d \tilde{\theta}'_{i \a} + i d \tilde{X}_{\a \dot{\g}} \bar{\theta}^{\dot{\g}}_{i}) \wedge ( d \tilde{\theta}'_{j \b} + i d \tilde{X}_{\b \dot{\delta}} \bar{\theta}^{\dot{\delta}}_{j})   + \tfrac{1}{2} Z^{ij} d \bar{\theta}^{\dot{\a}}_{i} \wedge d \bar{\theta}_{j \dot{\a}} \biggr\}, 
\eea
where we have redefined $\tilde{\theta}'_{i \a} = \tilde{\theta}_{i \a} - i \tilde{X}_{\a \dot{\b}} \bar{\theta}^{\dot{\b}}_{i}$. 

 The important observation now is that this fermionic T-dual action is related to the original action in a different choice of $\kappa$-symmetry gauge. So, instead of setting the fermionic coordinates that couple to the $S$-generators to zero at the beginning, if one relaxes the reality condition (\ref{real}) on the fermionic coordinates,  one can consider the following choice 
\be
\theta^{i \a}_{-} = 0 = \bar{\theta}^{\dot{\a}}_{+i}. 
\ee
This results in a complexification of the $AdS_5 \times S^5$ action and can be seen as the direct analogue of relaxing the Majorana condition on the spinors in (\ref{compspinor}). 

In this new complex $\kappa$-symmetry gauge, one can repeat the steps above and determine $j$. In this case a mixture of $Q$ and $\bar{S}$ generators are retained and the $\bar{Q}$ and $S$ parts are gauged away. Up to various field redefintions one discovers that the action in this complex $\kappa$-symmetry gauge and the action that results from doing a succession of bosonic and fermionic T-dualities are the same \cite{Beisert:2008iq}. This implies that the original $AdS_5 \times S^5$ action after bosonic and fermionic T-dualities has an equivalent superconformal $PSU(2,2|4)$ global symmetry group. In particular, this means that the corresponding Noether charges of the dual model should have their origin in the hidden charges of the original model, and vice versa. 

%Recall that in general sigma-model dualities do not respect manifest global symmetries. As an example, we can consider T-duality for an $S^2$ with metric $ds^2 = d \theta^2 + \sin^2 \theta d \phi^2$ and perform a T-duality to the $\phi$-direction to get the metric $ds^2 = d \theta^2 + \sin^{-2} \theta d \phi^2$. The two sigma-models are still classically equivalent, but the first model has $SO(3)$ symmetry and the second $SO(2)$. In this context, $AdS_n$ presents us with a special case, where T-duality along the translational directions and an inversion of the $AdS_n$ radial direction brings us back to $AdS_n$. This means that the original and dual models happen to have equivalent global Noether symmetries $SO(2, n-1)$ but realised on different variables. 

Since the two dual models are classically equivalent and share the same integrable structure, the local Noether charges of the dual model should be related to the hidden (non-local) charges of the dual model and vica versa. As was shown in \cite{Ricci:2007eq}, for T-duality on AdS spaces the Lax connection \cite{Bena:2003wd} can be expressed in terms of either the original or dual variables and thus, the charges in the two pictures are related. Now, by taking account of fermionic T-duality, one can show that the Lax connections of the original and dual sigma models may be regarded as equivalent. 

To see this one introduces a $\mathbb{Z}_4$-automorphism $\Omega$ of the superconformal algebra $\mathfrak{psu}(2,2|4)$ \cite{Beisert:2008iq}. This allows us to decompose the current as in (\ref{current}) for our original $\kappa$-symmetry gauge (where $S$-generators do not appear) 
\be
j = j_{P} + j_{D} + j_{R} + j_{Q} +j_{\bar{Q}}, 
\ee
and by taking into account the combined bosonic and fermionic T-dualities and various coordinate transformations \cite{Beisert:2008iq}, one can see that the action on the current is of the form:
\bea
\label{duality}
j_{P} &\mapsto& * j_{P}, \quad j_{D} \mapsto - j_{D}, \nn
j_{R_a} &\mapsto& - j_{R_a}, \quad j_{R_s} \mapsto j_{R_s}, \nn
j_{Q} &\mapsto& i j_{Q}, \quad j_{\bar{Q}} = \Omega(j_{\bar{Q}}), 
\eea
where $R_{s}$ and $R_{a}$ represent symmetric $R_{(ij)}$ and anti-symmetric $R_{[ij]}$ R-symmetry generators. 
Thus, starting from the family of flat currents or Lax connection in the original $\kappa$-symmetry gauge \cite{Beisert:2008iq}
\bea
\label{lax1}
j(z) &=& \tfrac{1}{4} (z+z^{-1})^2 j_P - \tfrac{1}{4}(z-z^{-1})^2 \Omega(j_{P}) - \tfrac{1}{4} (z^2-z^{-2}) * (j_P - \Omega(j_P)) \nn &+& \tfrac{1}{2} (z^2 +z^{-2}) j_{D} - \tfrac{1}{2} (z^2 - z^{-2}) * j_{D} \nn  &+& \tfrac{1}{2} (z+z^{-1}) (j_{Q} + j_{\bar{Q}}) - \tfrac{i}{2} (z-z^{-1}) (\Omega(j_{Q}) + \Omega(j_{\bar{Q}})), 
\eea
after applying the duality transformations (\ref{duality}), we obtain the dual flat current family,
\bea
\label{lax2}
\tilde{j}(z) &=& \tfrac{1}{4} (z+z^{-1})^2 *j_{P} - \tfrac{1}{4} (z-z^{-1})^2 * \Omega(j_P) - \tfrac{1}{4} (z^2-z^{-2}) (j_P - \Omega(j_P)) \nn &-& \tfrac{1}{2} (z^2 + z^{-2}) j_D + \tfrac{1}{2} (z^2-z^{-2}) * j_D\nn &+& \tfrac{i}{2} (z+z^{-1}) (j_{Q}-i \Omega(j_{\bar{Q}})) + \tfrac{1}{2} (z- z^{-1}) (\Omega(j_{Q}) + i j_{\bar{Q}}), 
\eea
which superficially appears to be a different Lax connection. However, it can be shown \cite{Beisert:2008iq} that the two Lax connections (\ref{lax1}) and (\ref{lax2}) are related by a $z$-dependent automorphism of the superconformal algebra, 
\be \tilde{j}(z) = \mathcal{U}_{z} (j(z)), 
\ee
where the automorphism acts on the generators $T$ of the superconformal algebra $\mathfrak{g}$ in the following way 
\be
T \mapsto \mathcal{U}_{z} (T) = {U}_{z} \Omega(T) U_{z}^{-1}, ~~\textrm{where}~~U_{z} = \left( \frac{z-z^{-1}}{z+z^{-1}}\right)^{i (B+D)}, 
\ee 
and $B$ generates a $U(1)$-automorphism, with the non-vanishing commutators being 
\be
[B, Q] = \tfrac{i}{2} Q, \quad [B,S] = - \tfrac{i}{2} S, \quad [B, \bar{Q} ] = - \tfrac{i}{2} \bar{Q}, \quad [B, \bar{S}] = \tfrac{i}{2} \bar{S}, 
\ee
and $\Omega(B) = - B$. 
This automorphism can, in principle, be used to obtain a map between the full set of conserved charges before and after the duality. 

\section{T-duality of $AdS_4 \times \mathbb{C} \textrm{P}^3$}
\label{sec:ads4cp3}
Unquestionably, one of the most interesting facets of fermionic T-duality has been the search for another example beyond $AdS_5 \times S^5$. If fermionic T-duality does indeed underlie hidden symmetries seen in scattering amplitudes, it is imperative that we find a separate manifestation of this symmetry. A promising place to look is ABJM theory \cite{ABJM}, a less supersymmetric setting in one dimension lower, which is AdS/CFT dual to the geometry $AdS_4 \times \mathbb{C} \textrm{P}^3$. From extensive work on scattering amplitudes in ABJM, both at tree level \cite{Agarwal:2008pu,Bargheer:2010hn, Lee:2010du, Huang:2010qy, Gang:2010gy} and loop-level \cite{Bianchi:2011fc, Bargheer:2012cp, Bianchi:2012cq, Brandhuber:2012un, Brandhuber:2012wy, Huang:2010qy}, we have gradually built up strong evidence for the existence of dual superconformal invariance, at least at weak-coupling. A noticeable example in support of this point of view is the agreement of the four-point two-loop amplitude \cite{Bianchi:2011dg,Chen:2011vv} with the two-loop Wilson loop \cite{Henn:2010ps}\footnote{Further investigations of Wilson loop/amplitude duality have also appeared in the literature \cite{Wiegandt:2011uu}.}.  

What is not clear at the moment is if these symmetries observed perturbatively persist at strong coupling. As the parallels to the symmetries observed in studies of $\mathcal{N}=4$ SYM are strong (also at the amplitude level \cite{Bianchi:2011aa}), it is reasonable to expect that fermionic T-duality could also play a role in a self-dual mapping of $AdS_4 \times \mathbb{C} \textrm{P}^3$. However, in contrast to $AdS_5 \times S^5$, the bosonic T-dualities along the Poincar\'e coordinates of $AdS_4$ will flip the chirality of the theory, and as fermionic T-duality preserves chirality, a more involved series of bosonic and compensating fermionic T-dualities will need to be considered if $AdS_4 \times \mathbb{C} \textrm{P}^3$ is to be self-dual. So at the moment the consensus is that a  recipe of three bosonic T-dualities along $AdS_4$, three internal $\mathbb{C} \textrm{P}^3$ T-dualities (to return to IIA) and 6 co
mpensating fermionic T-dualities are required. This possibility was first  suggested in \cite{Bargheer:2010hn} from studies of the superconformal algebra. The 6 compensating fermionic T-dualities are natural here as from the 12 Poincar\'e supersymmetries preserved by $AdS_4 \times \mathbb{C} \textrm{P}^3$, we can form 6 pairs and 6 commuting fermionic isometry directions. Indeed, this is analogous to $AdS_5 \times S^5$, where from 16 Poincar\'e supercharges, one picks out 8 commuting fermionic isometries. 

Despite the immediate hurdles, neglecting the chirality problem, early attempts have been made to combine bosonic T-duality along $AdS_4$ with fermionic T-duality. In \cite{Adam:2009kt} generalisations to models based on supercosets of the ortho-symplectic supergroup were considered. This class includes the OSp($6|4$) supercoset construction of $AdS_4 \times \mathbb{C} \textrm{P}^3$ \cite{Arutyunov:2008if, Stefanski:2008ik}. Within that context, it was shown for the supercoset construction with the $\kappa$-symmetry partially fixed, that the fermionic T-duality transformation is singular. The possibility was raised that the OSp($6|4$) supercoset, since it requires a partial $\kappa$-symmetry fixing which is not compatible with all string solutions, may not be sufficient and that the complete unfixed sigma-model derived in \cite{Gomis:2008jt} should be used. However, a follow-up study \cite{Grassi:2009yj} using an alternative $\kappa$-symmetry fixing of the complete $AdS_4 \times \mathbb{C} \textrm{P}^3$ superspace, which is suitable for studying regions of the theory not reached by the supercoset sigma-model, has concluded that it is not possible to T-dualise the fermionic sector of the superstring action on $AdS_4 \times \mathbb{C} \textrm{P}^3$. 

While the 3+6 recipe, i.e. 3 bosonic T-dualities along $AdS_4$ and 6 compensating fermionic T-dualities, which mimics one of the transformations of $AdS_5 \times S^5$, can be discounted, a later study of the superconformal algebra \cite{Bargheer:2010hn} suggested that 3+3+6 should be the correct set of T-dualities. This has the natural advantage that one returns to IIA and also has a close analogue in the second self-duality transformation of $AdS_5 \times S^5$ presented in \cite{Berkovits:2008ic}. Subsequently, two studies appeared \cite{Adam:2010hh,Bakhmatov:2010fp}, one \cite{Adam:2010hh} confining its attention to the OSp($6|4$) supercoset and the other \cite{Bakhmatov:2010fp} offering a complementary supergravity treatment. In both cases,  singularities were observed in the transformation. As a small positive development, it has been observed that the pp-wave limit of $AdS_4 \times \mathbb{C} \textrm{P}^3$ is self-dual \cite{Bakhmatov:2011aa}, so certainly in some sector of ABJM theory, we have a symmetry with respect to fermionic T-duality.  

Expectations that  the hidden symmetries of scattering amplitudes observed perturbatvely in ABJM will also be observed at strong-coupling via the AdS/CFT dual geometry have been fueled by numerous close analogies to $\mathcal{N} =4$ SYM and its dual geometry $AdS_5 \times S^5$. So, as pointed out in \cite{Adam:2010hh}, the singularities in the transformation may simply mean that dual superconformal symmetry exists only at weak-coupling and breaks down in the strongly-coupled regime where we are attempting to apply fermionic T-duality. A related possibility is that the symmetry exists, but it is not a fermionic T-duality that relates the dual superconformal symmetry to the ordinary superconformal symmetry and there is a more intricate relationship. 

 However, one further possibility  remains. It is known that the coset formulation does not describe the entire superstring and that, starting from the full $AdS_4 \times \mathbb{C} \textrm{P}^3$ sigma-model \cite{Gomis:2008jt},  the choice of $\kappa$-symmetry fixing may affect the outcome. Thus, a third possibility is that we have performed a potentially inconsistent truncation of the theory via the gauge-fixing of the $\kappa$-symmetry and the dual superconformal symmetry is not preserved. One could try to avoid this problem by simply working with fermionic T-duality transformations in the supergravity as has been done in \cite{Bakhmatov:2009be,Bakhmatov:2010fp,Bakhmatov:2011aa, OColgain:2012ca}.  Although, in moving to a supergravity treatment, one encounters another problem, notably an appropriate complexification for the $\mathbb{C} \textrm{P}^3$. Indeed, with a particular choice of complexified $\mathbb{C} \textrm{P}^3$, the supergravity treatment in \cite{Bakhmatov:2010fp} also encounters a singularity hinting that this may not be an artifact of $\kappa$-symmetry fixing. A more systematic treatment covering other possibilities for complexification would be welcome. 
 
\section{Generalisations}

In this section we review studies of fermionic Tduality in settings where there is no immediate connection to scattering amplitudes, or where the motivation is simply to develop a better formal understanding of aspects of this new version of T-duality. We begin by addressing geometries which exhibit a self-dual property under fermionic T-duality. 

\subsection{Self-dual geometries} 
Working with Green-Schwarz sigma-models, we have witnessed a few papers studying both critical and non-critical strings \cite{Adam:2009kt,Dekel:2011qw,Hao:2009hw}. As the calculations involved closely mirror those of section \ref{sec:ads5s5}, we omit technical details associated to T-duality. Building on earlier work \cite{Adam:2009kt}, in which it is shown that Green-Schwarz sigma-models based on supercosets PSU supergroups, such as $AdS_2 \times S^2$ and $AdS_3 \times S^3$, are self-dual, a general classification of fermionic T-duality symmetries of integrable Green-Schwarz sigma-models on Anti-de Sitter backgrounds with RR fluxes followed in  \cite{Dekel:2011qw}. 

%while supercosets of OSp supergroups, such as non-critical $AdS_2$ and $AdS_4$ models, and critical $AdS_4 \times \mathbb{C} \textrm{P}^3$ (discussed in section \ref{sec:ads4cp3}) are not. A little later, it was confirmed that the sigma model with $AdS_5 \times S^1$ background, which is the supercoset of SU($2,2|2$) supergroup, is self-dual \cite{Hao:2009hw}. %Furthermore, new integrable Green-Schwarz sigma-models with background $AdS_2 \times \mathbb{C} \textrm{P}^n$ are presented and show to be self-dual when taken as supercosets of SU supergroups, but not OSp supergroups \cite{Hao:2009hw}. 
%These claims about self-duality are certainly striking, since if true, they would present the first example where the dimensionality of AdS and internal spaces are different!\footnote{$AdS_4 \times \mathbb{C} \textrm{P}^3$ is a notable example.} However, the fact that the dilaton shifts is later pointed out in \cite{Dekel:2011qw}, so though classically self-dual, at the quantum level this is no longer true. 

The work of \cite{Dekel:2011qw} presents a general treatment for semi-symmetric backgrounds ($\mathbb{Z}_{4}$ supercoset spaces) and identifies criteria for a background to be self-dual. Denoting the superconformal algebra (SCA) $\mathfrak{g}$, one may further decompose the SCA according to a $\mathbb{Z}$-gradation with gradings $\pm 1, 0$, where the charges are assigned by a generator $U$. As an example of a familiar $\mathbb{Z}$-gradation, using the commutation relations of the SCA 
\bea
&&[P,Q] = 0, \quad [K,S] =0, \quad [P,S] \sim Q, \quad [K,Q] \sim S, \nn
&& [R,Q] \sim Q, \quad [R,S] \sim S, \nn
&&\{Q,Q \} \sim P, \quad \{ S, S \} \sim K, \nn
&& \{Q,S\} \sim D + L +R, 
\eea
we can decompose the SCA using the charge of the generators under the dilatation generator $D$ (Table 1). This gives a decomposition of the algebra of the form $\mathfrak{g} = \oplus_{i \in \mathbb{Z}} \mathfrak{g}_i $ with $[\mathfrak{g}_i, \mathfrak{g}_{j} ] \subset \mathfrak{g}_{i+j}$, and for the particular case of the gradation based on the dilatation generator, $-2 \leq i \leq 2$. In addition to this $\mathbb{Z}$-gradation the SCA may have others, such as a \textit{distinguished gradation} \cite{Frappat:1996pb}. By combining gradations \cite{Dekel:2011qw} one can identify various types of $\mathbb{Z}$-gradation of the SCA, such as for superalgebras that are a direct sum of two irreducible representations, we have the following:
\bea
\mathfrak{g}_{1} &=& (P, Q)_{1} \oplus (L, D, \hat{Q}, \hat{S}, R)_{0} \oplus (K, S)_{-1}, \nn
\mathfrak{g}_{2} &=& (P, Q^k, \hat{Q}_{k}, R_{k}^{~l})_{1} \oplus (L, D, Q^{k'}, \hat{Q}_{k} S_{k'}, \hat{S}^{k} R_{k}^{~l} R_{k'}^{~l'} )_{0} \oplus (K, S_k, \hat{S}^{k'}, R_{k}^{~l'} )_{-1}, \nn
\mathfrak{g}_3 &=& (Q, \hat{S})_1 \oplus (P, K, D, L, R)_0 \oplus (\hat{Q}, S)_{-1}, 
\eea
where the subscripts on the algebra $\mathfrak{g}$ simply label different types of $\mathbb{Z}$-gradation. What is common is that, appropriately normalised, we have the gradings $\pm 1, 0$. 

\begin{table}[h]
\begin{center}
\label{table:D}
\begin{tabular}{c|c|c|c|c}
$~~K~~$ & $~~S~~$ & $~~D, L, R~~$ & $~~Q~~$ & $~~P~~$ \\
\hline
-2 & -1 & 0 & 1 & 2 \\
\end{tabular}
\caption{The charge of the SCA generators under $D$.}
\end{center}
\end{table}

The gradation $\mathfrak{g}_1$ appeared in \cite{Berkovits:2008ic,Beisert:2008iq}, $\mathfrak{g}_2$ was mentioned in \cite{Berkovits:2008ic} where $R_{k}^{~l}$ correspond to internal bosonic T-dualities along the $S^5$ (the R-symmetry), whereas the last possibility is introduced in \cite{Dekel:2011qw} and implies that the geometry may be self-dual under T-duality along fermionic directions only. For each choice of gradation T-duality is performed along all the directions with charge $1$, which form an Abelian subalgebra. 

The background is then self-dual provided it satisfies a number of criteria \cite{Dekel:2011qw}: 
\begin{enumerate}
\item $\Omega(U) = -U$, where $\Omega$ is the $\mathbb{Z}_4$ autmorphism map. 

\item Rank($\kappa$-symmetry) $\geq \textrm{dim} (\mathfrak{g}_{o})/4$. 

\item The SCA's Killing-form vanishes. 
\end{enumerate}
Here, as stated earlier, $U$ is the generator whose charges determine the $\mathbb{Z}$-gradation, $\mathfrak{g}_{o}$ denotes the odd part of the algebra $\mathfrak{g}$ and the Killing-form is defined as the supertrace of every two generators in the adjoint representation \cite{Frappat:1996pb,VGKac}. The first condition ensures a non-singular coupling of fermionic coordinates. The second condition allows a particular representation of the supergroup that is used in the T-duality procedure, while the last condition guarantees that a non-trivial dilaton is not generated. 

The three backgrounds identified that are consistent with these conditions are $AdS_n \times S^n$, for $n=2, 3, 5$, all of which had been previously identified in the literature \cite{Adam:2009kt, Berkovits:2008ic, Beisert:2008iq}. Further examples, such as $AdS_n \times S^1, n=2, 3, 5$, $AdS_4 \times S^2$ and $AdS_2 \times S^4$, while classically self-dual (this possibility was first raised in \cite{Hao:2009hw}), fail to be self-dual at the quantum level as a non-trivial dilaton is generated. $AdS_4 \times \mathbb{C} \textrm{P}^3$ fails to satisfy the first of these conditions. 

From the supergravity perspective, a number of papers have also explored self-dual geometries. Beginning with \cite{Bakhmatov:2009be}, which also considers fermionic T-duality transformations on D1-branes, it has been shown that the maximally supersymmetric pp-wave in type IIB supergravity \cite{Blau:2001ne} is self-dual with respect to eight commuting fermionic isometries. This research thread was further picked up in \cite{Bakhmatov:2011aa} where it was shown that pp-waves in type IIA supergravity are self-dual in the same way. Since pp-waves  typically preserve sixteen standard supersymmetries, from which one has the freedom to construct eight commuting fermionic isometries, it was further conjectured that self-duality under fermionic T-duality is a symmetry of all pp-waves  \cite{Bakhmatov:2011aa}. 

Various aspects of self-duality in the geometry $AdS_3 \times S^3 \times CY_2$, where $CY_2$ is a Calabi-Yau two-fold, were discussed in \cite{OColgain:2012ca}\footnote{It is expected that, in line with the Green-Schwarz sigma-model analysis \cite{Adam:2009kt},  the geometry $AdS_2 \times S^2 \times CY_3$ is also self-dual.}. In addition to studying the effect of $S$-duality on self-dual geometries, \cite{OColgain:2012ca} presented a supergravity realisation of a self-duality transformation involving internal bosonic T-dualities along a complexified $S^3$, the possibility of which was first mentioned in \cite{Berkovits:2008ic}. As this is the only explicit calculation\footnote{A similar supergravity calculation was attempted in \cite{Bakhmatov:2010fp} though, as the background in question was $AdS_4 \times \mathbb{C} \textrm{P}^3$, singularities were encountered. } of this nature in the literature, we recap some of the details. 

Starting form the usual metric on $S^3$
\be
ds^2(S^3) = d \theta^2 + \sin^2 \theta ( d \phi^2 + \sin^2 \phi d \psi^2), 
\ee
one can complexify the sphere through the following coordinate transformations 
\bea 
\label{comp_coord}
w&=& \frac{i}{\sin \theta \sin \phi} e^{-i \psi}, \quad
x_1 = \frac{i \cos \theta}{\sin \theta \sin \phi} e^{-i \psi}, \quad
x_2 = \frac{i \cos \phi}{\sin \phi} e^{-i \psi}, 
\eea
to find a de Sitter metric
\be
\label{spheremetcomp}
ds^2(S^3_{\mathbb{C}})=\frac{-dw^2+dx_1^2+dx_2^2}{w^2}.
\ee
Having complexified the $S^3$ in this fashion we now have two commuting Killing directions and can perform two internal bosonic T-dualities. The fermionic T-dualities that will bring the geometry back to its original guise are then built from the Killing spinors $\eta$ invariant under these directions. To identify these directions, we utilise the spinorial Lie derivative \cite{Kosmann, FigueroaO'Farrill:1999va}
\be
{\cal L}_{K} \eta = K^{M} \nabla_{M} \eta + \frac{1}{8} d K_{MN} \G^{MN} \eta. 
\ee 
 After a small calculation one finds that both $K = \partial_{x^1}$ and $K = \partial_{x^2}$ lead to the same projection condition on the Killing spinors, thus uniquely determining the Killing spinors for the fermionic T-dualities \cite{OColgain:2012ca}. 

While analysis based on Green-Schwarz sigma-models and direct supergravity treatment should be regarded as equivalent, subtle differences can arise. For example, the Green-Schwarz sigma-model on $AdS_3 \times S^3$ \cite{Rahmfeld:1998zn,Pesando:1998wm,Park:1998un} is self-dual with respect to two bosonic T-dualities along $AdS_3$ and four fermionic T-dualities \cite{Adam:2009kt}. In $D=10$ where the related geometry is $AdS_3 \times S^3 \times T^4$, this combination of T-dualities brings the geometry back modulo the distinction that instead of being sourced by a three-form flux, it is sourced by a five-form flux \cite{OColgain:2012ca}. In other words, the original D1-D5 system is replaced by intersecting D3-branes and further T-dualities are required. 

%It is possible that a systematic supergravity treatment testing the self-duality properties of $AdS_4 \times \mathbb{C} \textrm{P}^3$ may offer some new insights. The real obstacle to be overcome is the choice of an appropriate complexification for $\mathbb{C} \textrm{P}^3$. 
 
\subsection{Related work} 
Unrelated to whether geometries exhibit a self-duality transformation incorporating fermionic T-duality, a small body of works studying some formal aspects of fermionic T-duality have appeared in the literature \cite{Chandia:2009yv,Fre:2009ki,Sfetsos:2010xa,Grassi:2011zf,ChangYoung:2011rs,Nikolic:2011ps,Nikolic:2012ih}. We now present brief summary of these papers. 

In \cite{Chandia:2009yv} both bosonic and fermionic T-duality in the context of pure spinor heterotic superstring are studied.  \cite{Fre:2009ki} attempts to treat fermionic T-duality in a background independent matter by defining a supersymmetric sigma-model which is globally invariant with respect to a super-duality group. In \cite{Sfetsos:2010xa} it is shown that fermionic T-duality, like its bosonic counterpart, can be viewed as a canonical transformation in phase space. \cite{Grassi:2011zf} explores extensions of fermionic T-duality beyond the classical approximation.  In \cite{ChangYoung:2011rs} a connection between fermionic T-duality and the Morita equivalence \cite{Rieffel:1998,Schwarz:1998qj} for noncommutative supertori is established. Noncommutativity in momenta can also be shown to arise as a consequence of fermionic T-duality \cite{Nikolic:2011ps}. Finally, aspects of Dirichlet boundary conditions in the context of fermionic T-duality are touched upon in \cite{Nikolic:2012ih}. 

\section{Outlook}
In contrast to its bosonic Abelian counterpart, fermionic T-duality is still relatively poorly understood. However, in light of the fact that it took a long time to understand how RR fluxes transformed under Abelian T-duality, we have witnessed swift progress in this direction driven along by many exciting developments in scattering amplitude research. To date, we have identified a handful of self-dual geometries, all of which are of the form $AdS_{n} \times S^{n}$, $n=2,3, 5$ \cite{Berkovits:2008ic, Beisert:2008iq, Adam:2009kt, Dekel:2011qw, OColgain:2012ca}, and only for $n=5$ do we have an interpretation in terms of symmetries of scattering amplitudes of $\mathcal{N}=4$ SYM  \cite{Berkovits:2008ic, Beisert:2008iq}. It would be interesting to understand what fermionic T-duality tells us about lower-dimensional SCFTs. 

Fermionic T-duality transformations applied to $AdS_4 \times \mathbb{C} \textrm{P}^3$ have encountered singularities \cite{Adam:2009kt, Grassi:2009yj, Adam:2010hh, Bakhmatov:2010fp}, and just on dimensionality grounds, if the geometry was self-dual, it is worth bearing in mind that this would be the first example where Anti-de Sitter and sphere factors differ in dimension! It remains to be seen if these singularities are an artifact of the $\kappa$-symmetry fixing or, alternatively in the supergravity, a choice of complexification for $\mathbb{C} \textrm{P}^3$. As pointed out in the text, there is substantial evidence for dual superconformal invariance from studies of scattering amplitudes in ABJM theory \cite{ABJM}.  Thus, it would constitute considerable progress if we could either resolve the singularities, or rule out fermionic T-duality in its current guise as the mechanism by which $AdS_4 \times \mathbb{C} \textrm{P}^3$ enjoys a self-dual property. This is an intriguing puzzle. 

Finally, fermionic T-duality has limited appeal as a solution generating technique since, as the Killing spinors get complexified, the resulting backgrounds are likely also to be complex. To date, no real solution has been found using this symmetry, but with the more general ansatz of  Godazgar \& Perry \cite{Godazgar:2010ph}, formally this is possible. More generally still, it is known that fermionic T-duality and S-duality do not commute, so an analogous ``fermionic U-duality" \cite{Hull:1994ys} group should be identified as it may play a role in more general transformations. One may also wonder if it is possible that timelike T-duality \cite{Hull:1998vg} combined with fermionic T-duality may be used to generate new real solutions without AdS factors. The challenge remains to generate a new solution using this symmetry of type II supergravity.

\section*{Acknowledgements} 
We are grateful to Ilya Bakhmatov, Marco Bianchi, Tristan McLoughlin and Patrick Meessen for discussion and comments. E \'O C is partially supported by the research grants MICINN-09-FPA2009-07122 and MEC-DGI-CSD2007-00042.

\end{document}